# KIC 8462852 Brightness Pattern Repeating Every 1600 Days

Bruce Gary[1] and Rafik Bourne[2]

[1] Hereford Arizona Observatory, Hereford, AZ 85615

[2] East Cannington, Perth, Western Australia 6107

The star KIC 8462852 (aka, "Tabby's Star" and "Boyajian's Star") has been described as "the most mysterious star in the cosmos." The *Kepler* mission telescope revealed this star to vary in brightness during 4 years in ways never before seen, and with no obvious periodicity (Boyajian *et al.*, 2016). *Kepler* is no longer able to observe this part of the sky (due to a hardware failure), so ground-based telescopes have to be relied upon for monitoring it in a search for any repeating pattern that might be used to understand the two categories of variation:

1) many brief (several-day) fades that reach depths of up to 22 %, and

2) a much longer timescale (> 4 years) variation that has not yet been proven to repeat, reaching a depth of ~ 3 %.

Attempts to explain the erratic short-timescale fades initially focused on the idea of swarms of comets with giant tails produced as they approached the star. This model, and the many that followed, were quickly deemed unsatisfactory, and the suggestion was made that alien megastructures, or Dyson spheres, were under construction and orbiting the star with an inclination that produced transits.

Observations with a small backyard observatory began in 2015 for the initial purpose of verifying the suggestion that KIC 8462852, hereafter KIC8462, was undergoing a gradual fade; as a bonus it was hoped that short-term dips in brightness would also be observed. By early May, 2017 observations showed that a long-term fade was indeed occurring.

On May 19 an observing team coordinated by Dr. Tabetha Boyajian observed an abrupt 2 % fade. An astronomer's telegram was issued, and this motivated many astronomers to obtain "priority" observing time on some of the world's best observatories for intense monitoring of KIC8462.

The purpose of this AAS Research Note is to report on recent brightness changes of KIC8462 that provide evidence of a repeating pattern, which will be a key step in making progress for understanding this mystery.

Montet and Simon (2016) re-analyzed *Kepler* data and showed that during the first 3.0 years KIC8462 faded with a linear rate of ~ 0.4 %/year, and then a U-shaped fade began that reached an additional fade level of ~ 2.4 %. Most of the short-timescale dips occurred at the "bottom" of this U-shaped fade. A Kepler hardware failure ended observations of KIC8462 in the middle of the U-shape.

The Hereford Arizona Observatory was used to produce the 2-year light curve shown in the figure. The "U-Shaped Fade Model" is an empirical equation (consisting of cosine terms for each half of the U-shape) that fits HAO data. The U-shape fade lasts ~ 1.0 year, the first half of which bears a striking resemblance to the *Kepler* fade identified by Montet and Simon (2016). Another similarity of the HAO light curve to the *Kepler* light curve is the presence of short-timescale dips during the last half of the U-shaped fade.

The two U-shaped fades are ~ 1600 days (4.4 years) apart. This is also the repeat interval for the main group of *Kepler* short-term dips and the ground-based ones observed this year. In the figure the lowest depths for the short-term dips are not shown because the purpose of this light curve is to emphasize the "out-of-transit" pattern.

KIC8462 exhibits a faster rate of brightening than the rate of fading that started one year ago. This asymmetry, and the steepness of the brightening during recovery, will constrain theories for what causes the yearlong fade. The U-shaped fade may have recovered by the end of October but this needs confirmation from observations in November. If these observations show that the brightening has ended, the gradual and linear fade that was present before the start of the present U-shaped fade is expected to resume in the near future.

The latest brightening phase was predicted in a private communication (RB to BG in July, 2017), stating that it would "start in September or October" (as described on October 02 at www.brucegary.net/ts4/). This was at a time when no one else was suggesting such an occurrence. We are predicting a repeat of the U-shaped fade and the short-term dips in 2021.

The dip structure is the subject of intense study by a team led by Dr. Boyajian; those observations have dip structure that is essentially identical to the ones observed at HAO. The Boyajian data has the added value of sampling several wavelengths, which promises to elucidate the particle size distribution of the dust that transits in front of KIC8462 that produces dip fades.

KIC8462 is a very complicated system, as evidenced by the many categories of brightness variability. Any model will necessarily have to be similarly complicated. This translates to a greater need for observations lasting many years, and especially during our predicted repeat of the U-shaped and dip fades in 2021.

As models are developed for natural structures that can account for the observed fades (such as ring systems and comas) there will be a parallel effort needed for the development of the physical mechanisms that can produce such structures. Notable progress on this has been reported by Neslusan and Budaj (2016). There will surely be many more publications addressing this "most mysterious star in the cosmos."

REFERENCES

Boyajian, T. S., LaCourse, D. M., Rappaport, S. A., *et al.*, 2016, *MNRAS*, **457**, 3988

Montet, B. T. & J. D. Simon, 2016, *ApJ*, **830**, L39

Neslusan, L. and J. Budaj, 2017, *A&A*, **600**, A86

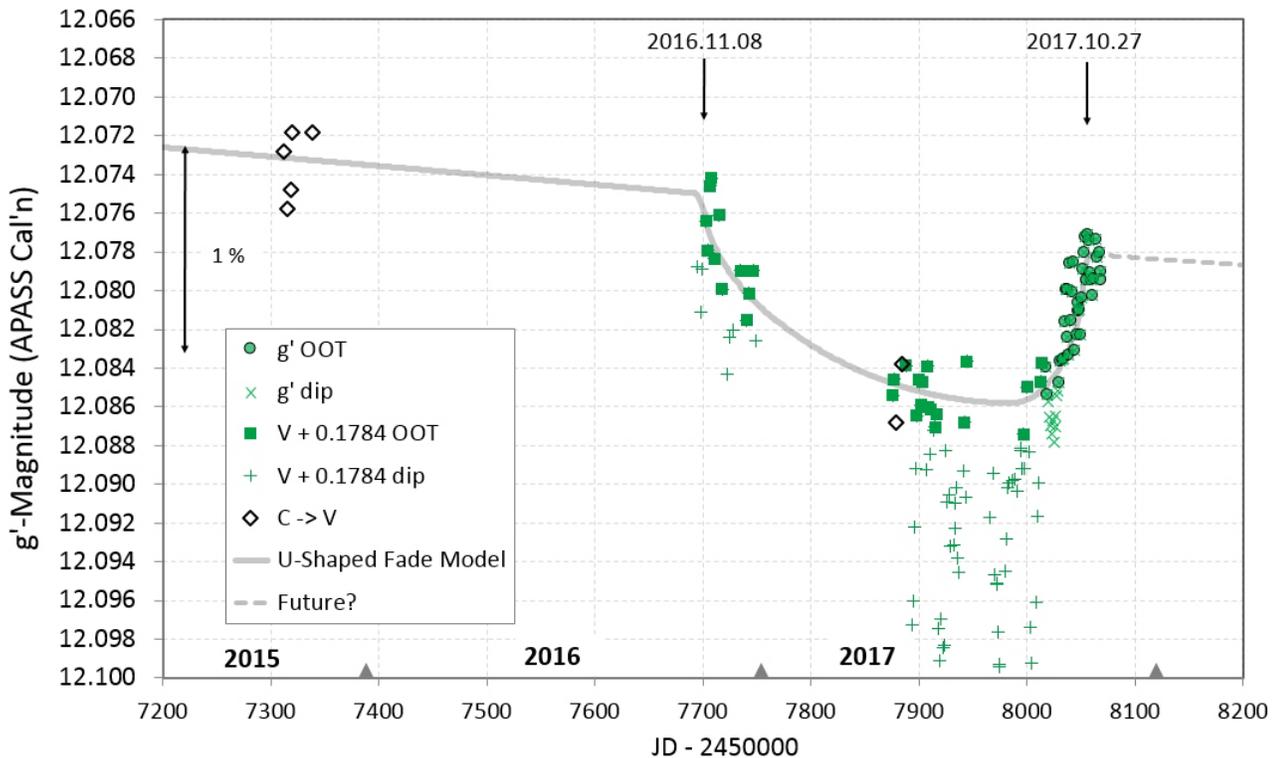

**Figure 1.** KIC8462 g'-magnitude versus date using filled-symbols for dates identified as "out-of-transit" and + or x symbols for dates identified as "dip in progress." The "U-Shaped Fade Model" is an empirical equation that fits HAO data. V and C filter observations have been converted to g'-band magnitude.